\documentclass{emulateapj}
\usepackage{natbib}
\usepackage{epsfig}
\usepackage{graphicx}
\usepackage{subfigure}
\usepackage{float}
\usepackage{amsmath}
\usepackage{color}
\usepackage{amssymb}
\usepackage{amsfonts}
\usepackage{apjfonts}
\usepackage[colorlinks,linkcolor=blue,anchorcolor=green,citecolor=blue]{hyperref}
\shorttitle{A Further Test of LIV from the Rest-Frame Spectral Lags of GRBs}
\shortauthors{Wei \& Wu}
\begin{document}

\title{A Further Test of Lorentz Violation from the Rest-Frame Spectral Lags of Gamma-Ray Bursts}

\author{Jun-Jie Wei\altaffilmark{1} and Xue-Feng Wu\altaffilmark{1,2}}

\affil{$^1$Purple Mountain Observatory, Chinese Academy of Sciences, Nanjing 210008, China; jjwei@pmo.ac.cn, xfwu@pmo.ac.cn \\
$^2$Joint Center for Particle, Nuclear Physics and Cosmology, Nanjing University-Purple Mountain Observatory, Nanjing 210008, China}

\begin{abstract}
Lorentz invariance violation (LIV) can manifest itself by an energy-dependent vacuum dispersion of light,
which leads to arrival-time differences of photons with different energies originating from the same astronomical source.
The spectral lags of gamma-ray bursts (GRBs) have been widely used to investigate the possible LIV effect.
However, all current investigations used lags extracted in the observer frame only. In this work, we present,
for the first time, an analysis of the LIV effect and its redshift dependence in the cosmological rest frame.
Using a sample of 56 GRBs with known redshifts, we obtain a robust limit on LIV by fitting their rest-frame
spectral lag data both using a maximization of the likelihood function and a minimum $\chi^{2}$ statistic.
Our analysis indicates that there is no evidence of LIV. Additionally, we test the LIV in different redshift ranges
by dividing the full sample into four redshift bins. We also find no evidence for the redshift variation of the LIV effect.
\end{abstract}


\section{Introduction}
\label{sec:intro}
Lorentz invariance is a fundamental symmetry of space-time in modern physics.
However, many Quantum Gravity (QG) theories, which attempt to unify general relativity
and quantum mechanics, predict the existence of deviations from Lorentz symmetry
at the Planck energy scale ($E_{\rm Pl}=\sqrt{\hbar c^{5}/G}\simeq1.22\times10^{19}$ GeV;
\citealt{1967PhRv..159.1106P,1989PhRvD..39..683K,1991NuPhB.359..545K,1995PhRvD..51.3923K,2005LRR.....8....5M,2005hep.ph....6054B,2013LRR....16....5A,2014RPPh...77f2901T}).
As a consequence of Lorentz invariance violation (LIV), the speed of light in a vacuum
would have an energy dependence, the so-called vacuum dispersion \citep{1997IJMPA..12..607A,2008ApJ...689L...1K,2008PhLB..665..412E,2011IJMPA..26.2243E}.
The energy scale for LIV, $E_{\rm QG}$, could therefore be constrained by comparing
the arrival-time differences of photons with different energies emitted simultaneously
from the same source \citep{1998Natur.393..763A,2013APh....43...50E}.

Because of the high-energy extent of their emission, their large cosmological distances, and
their fast variabilities, gamma-ray bursts (GRBs) have been deemed as the most promising sources
for searching for LIV-induced vacuum dispersion \citep{1998Natur.393..763A,2007NatPh...3...87J,2009PhRvD..80h4017A,2013LRR....16....5A,2016JCAP...08..031W}.
To date, various limits on LIV have been obtained by studying the dispersion of light in observations
of individual GRBs or a large sample of GRBs (\citealt{1998Natur.393..763A,
1999PhRvD..59k6008C,1999PhRvL..82.4964S,2004ApJ...611L..77B,2005PhLB..625...13P,
2006PhLB..643...81K,2008JCAP...01..031J,2009Sci...323.1688A,2009Natur.462..331A,
2009PhRvD..80k6005X,2010APh....33..312S,2012APh....36...47C,2012PhRvL.108w1103N,
2013PhRvL.110t1601K,2013PhRvD..87l2001V,2015APh....61..108Z,2016APh....82...72X},
see also \citealt{2011RvMP...83...11K,2013CQGra..30m3001L} and summary constraints for LIV therein).
Although these limits on LIV have reached high precision, most were obtained by relying on
the rough time lag of a single highest-energy photon. Performing a search for LIV
using the true time lags of high-quality and high-energy light curves in different energy multi-photon bands
is therefore crucial. Furthermore, the method of the arrival-time difference used for testing LIV is
tempered by our ignorance concerning the intrinsic time delay that depends on the unknown emission mechanism of GRBs.
Most previous studies concentrated on the time delay induced by LIV while neglecting the intrinsic time delay,
which would impact the reliability of the resulting constraints on LIV.
Most recently, however, \cite{2017ApJ...834L..13W,2017ApJ...842..115W} provided some solutions to disentangle
the intrinsic time delay problem. They first proposed that the only burst GRB 160625B so far with
a well-defined transition from positive to negative spectral lags\footnote{The spectral lag is defined as
the arrival time difference between high- and low-energy photons and is considered to be positive when high-energy
photons precede low-energy photons.} provides a good opportunity to distinguish
the possible LIV effect from any source-intrinsic time delay in the emission of photons of different energy bands.
By fitting the true multi-photon spectral-lag data of GRB 160625B, they obtained both a reasonable formulation of
the intrinsic energy-dependent time delay and robust limits on the QG energy scale and the Lorentz-violating coefficients
of the Standard-Model Extension.

In order to identify an effect as radical as Lorentz violation, statistical and possible systematic uncertainties must be
minimized. For this purpose, \cite{2000ApJ...535..139E,2003A&A...402..409E,2006APh....25..402E} developed a method for
analyzing a statistical sample of 35 GRBs with different redshifts, rather than a single source. For each GRB, \cite{2006APh....25..402E}
looked for the spectral time-lag in the light curves recorded in the selected observer-frame energy bands
25--55 and 115--320 keV. This technique has the advantage that it can extract the spectral lags of broad light curves in different energy multi-photon bands.
In their analysis, the observed time lag was formulated in terms of linear regression
where the slope corresponds to the LIV effect and the intercept denotes the intrinsic time delay at the source.
Assuming the concordance $\Lambda$CDM model as the background cosmology, they found a weak evidence for LIV and obtained a
robust lower limit of $E_{\rm QG}\geq1.4\times10^{16}$ GeV \citep{2006APh....25..402E}. Subsequently, \cite{2009CQGra..26l5007B,2015ApJ...808...78P}
applied this procedure to different cosmological models, finding the conclusion is independent of the background cosmology.
We note that the spectral lags of GRBs used for testing LIV were extracted between
two fixed energy bands in the observer frame \citep{2006APh....25..402E}. Due to the redshift dependence of GRBs, however,
these two energy bands can correspond to a different pair of energy bands in the rest frame~\citep{2012MNRAS.419..614U},
thus potentially introducing an energy dependence to the extracted spectral lag and/or an extra uncertainty to the resulting constraints on LIV.
\cite{2012MNRAS.419..614U} studied the correlation between observer-frame lags and rest-frame lags for the same sample of 31 GRBs
(see Figure~3 of \cite{2012MNRAS.419..614U}). They showed that there is a large scatter in this correlation,
implying the observer-frame lag does not directly represent the rest-frame lag.
In other words, the observer-frame lags would be strongly biased since they simply recorded different pairs of the intrinsic light curves.
This can be resolved by choosing two appropriate energy bands fixed in the rest frame and estimating the observed lag
for two projected energy bands by the relation $E_{\rm obs}=E_{\rm rest}/(1+z)$, where $E_{\rm obs}$ and $E_{\rm rest}$
are the photon energy measured in the observer and the rest frame, respectively.
It actually means that the energy-dependent effect can be removed in this way.
However, no one has made systematical research on the LIV effect using the spectral lags extracted in the rest frame.

Recently, \cite{2015MNRAS.446.1129B} investigated the rest-frame spectral lags of
the complete sample of 56 bright GRBs observed by \emph{Swift}/Burst Alert Telescope (BAT).
For each GRB, they extracted mask-weighted, background-subtracted light curves for
two observer-frame energy bands corresponding to the fixed rest-frame energy bands 100--150 and 200--250 keV.
These two particular rest-frame energy bands were selected so that after transforming to the observer frame
(i.e., $[100$--$150]/(1+z)$ and $[200$--$250]/(1+z)$ keV) they still lie in the energy range of
the BAT instrument ($\sim[15$--$200]$ keV). For each light-curve pairs,
they estimated the time lag and the associated uncertainty through a discrete cross-correlation function
fitted with an asymmetric Gaussian function \citep{2015MNRAS.446.1129B}.

In this work, we make use of the rest-frame spectral lags from 56 \emph{Swift} GRBs presented in
\cite{2015MNRAS.446.1129B} for the first time to study the possible LIV effect.\footnote{
Note that the redshift dependent photon energy difference of the \emph{Swift} data is in the range of several tens of keV,
while the energy difference of the \emph{Fermi} data is wider, which can reach several hundreds or thousands of keV.
However, since the difference in the arrival time delay between a higher and a lower energy photon is proportional to
the difference in the photon energies \citep{2017NatAs...1E.139A,2017arXiv170702413A},
the spectral lag of the \emph{Swift} data is smaller than that of the \emph{Fermi} data.
Therefore, the constraint on $E_{\rm QG}$ obtained from the \emph{Swift} data would be
the same order of magnitude as that of the \emph{Fermi} data (see Equation~\ref{eq:tLIV}).
That is, the resulting constraint could not benefit too much from using \emph{Fermi} data.}
Compared with previous works, which relied on using the observer-frame lags, our present work
can remove the problem associated with the energy dependence of the extracted spectral lags,
and then obtain reliable constraints on LIV.
On the other hand, since the GRBs in the sample of \cite{2015MNRAS.446.1129B} cover
a wide redshift range $z\in[0.35$--$5.47]$, they also enable us to test the LIV in different redshift ranges
for the first time. To test whether the LIV effect varies with the cosmological redshift, in this work
we separate the sample of 56 GRBs into four redshift bins. That is, we sort the GRBs by their redshift measurements,
and divide them into four groups with redshifts from low to high, each group containing 14 GRBs.
We then perform the linear fit to each group and calculate the mean redshift, and check whether the values of
the linear terms that encoded the LIV effect evolve with redshift.

The outline of this paper is as follows. In Section~\ref{sec:formalism}, we give an overview of the LIV formalism.
In Section~\ref{sec:data}, we describe the sample at our disposal and our method of analysis.
The resulting constraints on LIV and its redshift dependence from the rest-frame spectral lags of GRBs are presented
in Section~\ref{sec:results}. Finally, a brief discussion and conclusion are drawn in Section~\ref{sec:summary}.

\section{Formalism}
\label{sec:formalism}

As mentioned above, the speed of light would become energy-dependent in a vacuum due to the LIV effect.
The modified dispersion relation of photons can be described using the leading term of the Taylor series expansion
in the form
\begin{equation}
E^{2}\simeq p^{2}c^{2}\left[1-s_{\pm}\left(\frac{pc}{E_{{\rm QG},n}}\right)^{n}\right]\;,
\label{eq:dispersion}
\end{equation}
which corresponds to a modified photon velocity
\begin{equation}
v(E)=\frac{\partial E}{\partial p}\approx c\left[1-s_{\pm}\frac{n+1}{2}\left(\frac{E}{E_{{\rm QG},n}}\right)^{n}\right]\;,
\end{equation}
where $E_{\rm QG}$ represents the QG energy scale, the $n$th-order expansion of the leading term
stands for linear ($n=1$) or quadratic ($n=2$) energy dependence, and $s_{\pm}=\pm1$ is the ``sign of LIV''
\citep{2001PhRvD..64c6005A}. $s_{\pm}=+1$ ($s_{\pm}=-1$) corresponds to a decrease (an increase) in photon speed
with an increasing photon energy. For the sake of being comparable with the results of \cite{2006APh....25..402E}
and because the data used in this study are not sensitive to higher order terms, here we only consider the $n=1$ case,
and moreover we assume the sign parameter $s_{\pm}=-1$ in the formula.
Photons with higher energies would travel faster than those with lower energies in the case of $s_{\pm}=-1$,
which predicts a positive spectral lag due to LIV, i.e., $\Delta t_{\rm LIV}>0$.

Because of the spectral dispersion, two photons with different observer-frame energies ($E_{h}>E_{l}$)
arising from the same source would arrive at the observer with a time delay.
Taking into consideration the cosmological expansion, the LIV-induced time delay can be expressed by
\citep{2008JCAP...01..031J,2015APh....61..108Z}
\begin{equation}
\begin{aligned}
\Delta t_{\rm LIV}&=\frac{E_{h}-E_{l}}{H_{0}E_{\rm QG}}\int_{0}^{z}\frac{(1+z'){\rm d}z'}{h(z')}\\
                  &=\frac{E'_{h}/(1+z)-E'_{l}/(1+z)}{H_{0}E_{\rm QG}}\int_{0}^{z}\frac{(1+z'){\rm d}z'}{h(z')}\;,
\label{eq:tLIV}
\end{aligned}
\end{equation}
where $E'_{h}=E_{h}(1+z)$ and $E'_{l}=E_{l}(1+z)$ are the rest-frame energies, respectively.
Also, $h(z)=\sqrt{\Omega_{\rm m}(1+z)^{3}+\Omega_{\Lambda}}$ is the dimensionless Hubble expansion rate at $z$,
where the standard flat $\Lambda$CDM model with parameters $H_{0}=67.8$ km $\rm s^{-1}$ $\rm Mpc^{-1}$, $\Omega_{\rm m}=0.308$,
and $\Omega_{\Lambda}=1-\Omega_{\rm m}$ is adopted \citep{2016A&A...594A..13P}.

Due to the fact that the LIV-induced time delay $\Delta t_{\rm LIV}$ is likely to be accompanied by an unknown intrinsic time lag
caused by unknown properties of the source \citep{2006APh....25..402E,2009CQGra..26l5007B}, we take this possibility into account
by fitting the observed time lag with the inclusion of a term $\langle b \rangle$ specified in the rest-frame of the source,
as \cite{2006APh....25..402E} did in their treatment. Therefore, the observed arrival time delays consist of two terms
\begin{equation}
 \Delta t_{\rm obs}= \Delta t_{\rm LIV}+\langle b \rangle(1+z)\;,
 \label{eq:tobs}
\end{equation}
reflecting the possible LIV effect and intrinsic source effect, respectively.
Then we re-express Equation~(\ref{eq:tobs}) as a linear function in the form:
\begin{equation}\label{eq:fit}
  \frac{ \Delta t_{\rm obs}}{1+z}=a_{\rm LIV}K+\langle b \rangle,
\end{equation}
where
\begin{equation}
  K=\frac{1}{(1+z)^2}\int_{0}^{z}
  \frac{(1+z') {\rm d}z'}{h(z')}
  \label{eq:K}
\end{equation}
is a function of the redshift, and
\begin{equation}
 a_{\rm LIV}=\frac{E'_{h}-E'_{l}}{H_0E_{\rm QG}}
 \label{eq:aLIV}
\end{equation}
is the slope in $K$ which is related to the scale of Lorentz violation, whereas the intercept $\langle b \rangle$ denotes
the average effect of the intrinsic time lags of different pulses internal to the GRB. Note that intrinsic time lags apply
most strictly between individual pulses internal to GRBs (see \citealt{2003ApJ...594..385K,2005ApJ...627..324N}).
As shown by \cite{2004ApJ...610..361H} and \cite{2012Ap&SS.339..123Z}, different pulses inside the same GRB can have different intrinsic lags.
We adopt the view that using a single time lag $\langle b \rangle$ for an entire GRB would be sufficient to demonstrate the essential point
\citep{2006APh....25..402E,2009CQGra..26l5007B,2015ApJ...808...78P}, but this should be understood to be a statistical average over
the different lags of different pulses internal to the GRB.

Additionally, it should be pointed out that the intrinsic time lags are not generally a single
time offset between GRB light curves in different energy bands. As described by \cite{2002ApJ...579..386N},
the time lag is a cross correlation between light curves of the same pulse at two different energy bands.
This cross correlation is not caused, typically, by a single time offset true at all times during the pulse,
but rather a variable time offset that increases as the pulse progresses. This has been shown in some detail
by \cite{2000ApJ...544..805N} for GRB 930214C, and more generally by \cite{2009ApJ...705..372H}. As such,
the total measured time offset is typically zero at the beginning of each pulse (the Pulse Start Conjecture,
\citealt{2000ApJ...544..805N,2012MNRAS.419.1650N,2009ApJ...705..372H}), but may even be on the order of the duration of the pulse
near the temporal conclusion of the pulse. It would be more accurate to consider pulses as starting at the
same time at all energies but being increasingly stretched out at lower energies, typically.
Therefore, any LIV-induced time lag would act in a fundamentally different manner than an intrinsic time lag,
as measured. LIV-induced time lags would really be a single time offset, on the average, between GRB photons
at different energies. LIV-induced time lags between energy bands would be the same at all times in the
light curves of GRB pulses, whereas intrinsic time lags would differ. In summary, the intrinsic time lags
are attributes of the later parts of GRB pulses, here we simply use a single average lag $\langle b \rangle$
for an entire GRB.

\begin{table}[!h]
\caption{Data on spectral time lags for 56 GRBs with known redshifts. The values of $z$, $\Delta t_{\rm obs}$,
and the corresponding left ($\sigma_{\rm l}$) and right ($\sigma_{\rm r}$) uncertainties of
$\Delta t_{\rm obs}$ are collected from \cite{2015MNRAS.446.1129B}.}
\begin{center}
\begin{small}
\begin{tabular}{lllll}
\hline
GRB name      &  $z$    & $\Delta t_{\rm obs}$ (ms)  &  $\sigma_{\rm l}$ (ms) & $\sigma_{\rm r}$ (ms)\\
\hline
050318	&	1.44	&	-13.66	&	184.88	&	218.76	\\
050401	&	2.90	&	285.19	&	59.05	&	59.14	\\
050525A	&	0.61	&	54.72	&	25.42	&	25.59	\\
050802	&	1.71	&	555.80	&	386.11	&	395.90	\\
050922C	&	2.20	&	162.52	&	74.74	&	79.50	\\
060206	&	4.05	&	252.40	&	85.65	&	88.18	\\
060210	&	3.91	&	349.99	&	233.64	&	237.12	\\
060306	&	1.55	&	42.56	&	51.17	&	53.73	\\
060814	&	1.92	&	-100.01	&	138.04	&	138.73	\\
060908	&	1.88	&	230.04	&	169.95	&	175.42	\\
060912A	&	0.94	&	-7.09	&	82.58	&	83.49	\\
060927	&	5.47	&	14.26	&	111.90	&	111.69	\\
061007	&	1.26	&	27.05	&	25.42	&	26.88	\\
061021	&	0.35	&	-603.94	&	416.22	&	403.94	\\
061121	&	1.31	&	28.36	&	20.02	&	20.25	\\
061222A	&	2.09	&	6.07	&	145.67	&	139.01	\\
070306	&	1.50	&	-213.78	&	290.08	&	281.92	\\
070521	&	1.35	&	40.20	&	39.51	&	39.07	\\
071020	&	2.15	&	48.47	&	10.70	&	10.24	\\
071117	&	1.33	&	258.54	&	41.21	&	42.58	\\
080319B	&	0.94	&	30.29	&	21.67	&	19.18	\\
080319C	&	1.95	&	217.82	&	168.48	&	171.20	\\
080413B	&	1.10	&	96.00	&	61.91	&	59.56	\\
080430	&	0.77	&	44.04	&	564.87	&	634.35	\\
080603B	&	2.69	&	-43.59	&	67.38	&	63.01	\\
080605	&	1.64	&	53.65	&	36.46	&	37.38	\\
080607	&	3.04	&	90.99	&	91.44	&	101.78	\\
080721	&	2.59	&	-158.16	&	162.73	&	149.69	\\
080804	&	2.20	&	-347.40	&	618.25	&	623.99	\\
080916A	&	0.69	&	599.82	&	288.57	&	290.73	\\
081121	&	2.51	&	-10.41	&	245.62	&	266.41	\\
081203A	&	2.10	&	-39.23	&	198.37	&	175.09	\\
081221	&	2.26	&	99.44	&	77.55	&	80.56	\\
081222	&	2.77	&	129.02	&	81.04	&	86.36	\\
090102	&	1.55	&	522.53	&	278.44	&	304.17	\\
090201	&	2.10	&	-56.92	&	175.92	&	176.01	\\
090424	&	0.54	&	18.62	&	47.22	&	50.44	\\
090709A	&	1.80	&	-31.00	&	68.71	&	71.05	\\
090715B	&	3.00	&	70.66	&	304.24	&	385.39	\\
090812	&	2.45	&	168.71	&	338.84	&	343.29	\\
090926B	&	1.24	&	1031.73	&	861.13	&	887.57	\\
091018	&	0.97	&	163.65	&	147.37	&	149.05	\\
091020	&	1.71	&	-78.58	&	282.06	&	290.03	\\
091127	&	0.49	&	157.64	&	194.65	&	192.49	\\
091208B	&	1.06	&	84.20	&	31.61	&	31.60	\\
100615A	&	1.40	&	162.03	&	106.60	&	108.27	\\
100621A	&	0.54	&	924.74	&	727.39	&	677.68	\\
100728B	&	2.11	&	-115.00	&	456.44	&	406.26	\\
110205A	&	2.22	&	-125.63	&	136.21	&	144.66	\\
110503A	&	1.61	&	46.77	&	82.15	&	85.65	\\
051221A	&	0.55	&	-1.85	&	2.32	&	2.47	\\
070714B	&	0.92	&	5.58	&	35.01	&	31.56	\\
090510	&	0.90	&	-7.99	&	8.40	&	8.63	\\
101219A	&	0.72	&	-0.02	&	21.77	&	22.42	\\
111117A	&	1.30	&	3.24	&	10.70	&	10.10	\\
130603B	&	0.36	&	-3.44	&	5.58	&	7.27	\\
\hline
\end{tabular}
\label{tab:lag}
\end{small}
\end{center}
\end{table}

\section{Observational data and methodology}
\label{sec:data}

Unlike previous analyses which used the observer-frame spectral lags, we first take advantage of
the rest-frame lags of 56 \emph{Swift} GRBs presented in \cite{2015MNRAS.446.1129B}
to study the LIV effect and its redshift dependence. This complete sample
has redshifts ranging from 0.35 (GRB 061021) to 5.47 (GRB 060927), with a mean redshift of $\sim1.73$.
By selecting two appropriate energy bands in the observer frame (based on the redshift of each GRB,
i.e., $[100$--$150]/(1+z)$ and $[200$--$250]/(1+z)$ keV),
\cite{2015MNRAS.446.1129B} extracted light curves for the chosen rest-frame energy bands
100--150 and 200--250 keV. Note that the energy gap between the mid-points of the two rest-frame energy
bands is fixed at 100 keV, whereas in the observer frame, as expected, the energy gap varies depending on
the redshift of each GRB. For example, the gap is 74 keV in GRB 061021 and it is 16 keV in GRB 060927.
This is in contrast to the spectral-lag extractions performed in the observer frame where the energy gap is treated
as a constant \citep{2012MNRAS.419..614U}.

The spectral lags for two observer-frame energy bands corresponding to the fixed rest-frame energy bands
100--150 and 200--250 keV from 56 GRBs were carefully computed by \cite{2015MNRAS.446.1129B}.
This complete sample is listed in Table~\ref{tab:lag}, which includes the following information for each GRB:
(1) its name; (2) the redshift; (3) the observed spectral time lags $\Delta t_{\rm obs}$; and the corresponding uncertainties of
$\Delta t_{\rm obs}$, including (4) the left uncertainty $\sigma_{\rm l}$, (5) the right uncertainty $\sigma_{\rm r}$.

\begin{figure}
\centerline{\includegraphics[angle=0,width=0.55\textwidth]{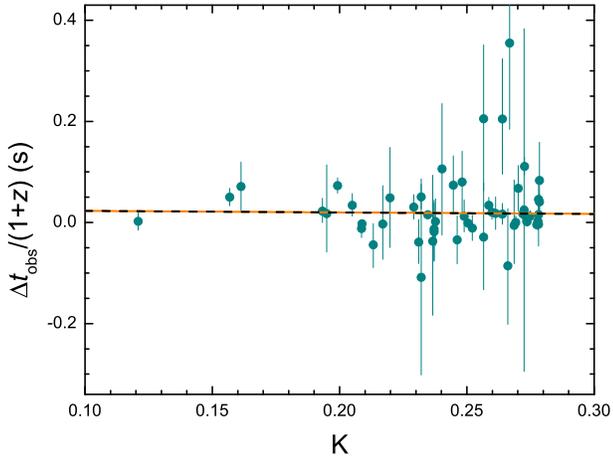}}
\vskip-0.1in
\caption{Dependence of the re-scaled spectral time lag $\Delta t_{\rm obs}/(1+z)$ on the variable $K(z)$.
The solid (dashed) line corresponds to the best-fit from the MLE (reduced $\chi^{2}$) method.}
\label{fig1}
\end{figure}

In order to probe the energy dependence of the speed of light that might be induced by the LIV effect,
following the treatment of \cite{2006APh....25..402E},
we perform a linear lit to the $\Delta t_{\rm obs}/(1+z)$ versus $K(z)$ data (see Figure~\ref{fig1}).
Since the data are quite scattered, we introduce a quantify, $\sigma_{\rm int}$, to characterize the intrinsic uncertainty in GRB spectral lag.
To find the best-fit coefficients $a_{\rm LIV}$, $\langle b \rangle$ and the intrinsic scatter $\sigma_{\rm int}$, we adopt the method of maximum likelihood estimation
(MLE; \citealt{2005physics..11182D,2013ApJ...772...43W,2015AJ....149..102W}).
The joint likelihood function for the coefficients $a_{\rm LIV}$, $\langle b \rangle$ and the intrinsic scatter $\sigma_{\rm int}$ is
\begin{equation}\label{likelihood}
\begin{aligned}
\mathcal{L}(a_{\rm LIV},\; \langle b \rangle,\; \sigma_{\rm int}) \propto &\prod_{i}
\frac{1}{\sqrt{\sigma_{\rm int}^{2}+\left(\frac{\sigma_{\Delta t_{i}}}{1+z_{i}}\right)^{2}}}\\
&\times\exp\left[-\,\frac{\left(\frac{\Delta t_{i}}{1+z_{i}}-a_{\rm LIV}K_{i}-\langle b \rangle\right)^{2}}{2\left(\sigma_{\rm int}^{2}+\left(\frac{\sigma_{\Delta t_{i}}}{1+z_{i}}\right)^{2}\right)}\right]\;,
      \end{aligned}
\end{equation}
where $\sigma_{\Delta t_{i}}=\left(\sigma_{{\rm l},i}+\sigma_{{\rm r},i}\right)/2$ is the observational uncertainty.

\section{Results}
\label{sec:results}

\subsection{Test of the LIV effect}

\begin{figure}
\centering
\vskip-0.1in
\begin{tabular}{c}
\includegraphics[keepaspectratio,clip,width=0.5\textwidth]{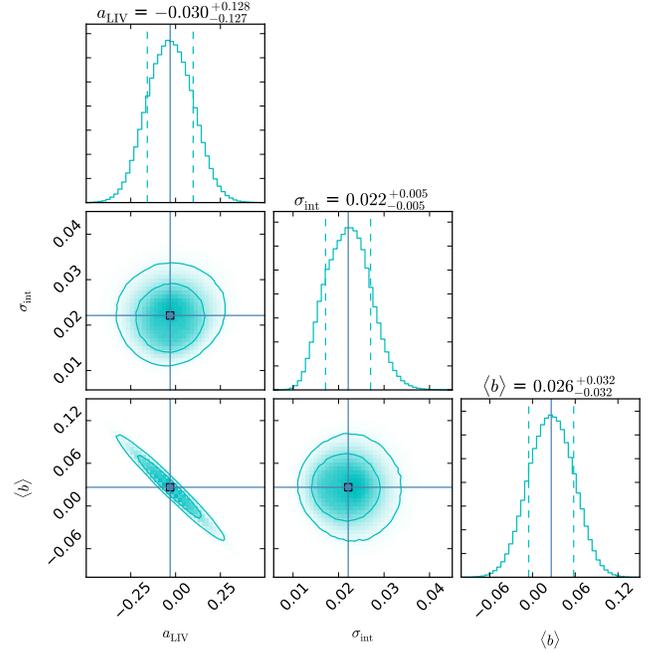} \\{(a) \;The maximum likelihood estimation method}\\
\includegraphics[keepaspectratio,clip,width=0.4\textwidth]{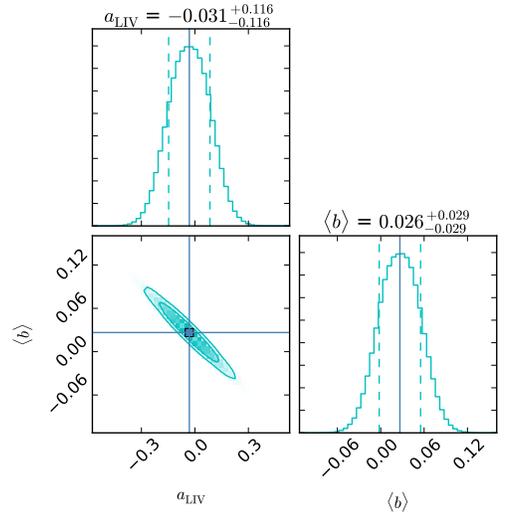} \\{(b) \;The reduced $\chi^{2}$ method}
\end{tabular}
\caption{(a): 1D marginalized distributions and 2D regions with the $1\sigma$ and $2\sigma$ contours of
the LIV parameters. The vertical solid lines represent
the best-fits, and the vertical dashed lines enclose the 68\% credible region.
Made with triangle.py from \cite{2013PASP..125..306F}. The fitting method employed is MLE.
(b): Same as panel (a), but now with the reduced $\chi^{2}$ method.}
\label{fig2}
\end{figure}

Using the MLE method, we get the resulting constraints on $a_{\rm LIV}$, $\langle b \rangle$, and $\sigma_{\rm int}$,
which are displayed in Figure~\ref{fig2}(a). These contours show that at the $1\sigma$ confidence level, the best-fit parameters are
$a_{\rm LIV}=-0.030^{+0.128}_{-0.127}$ and $\langle b \rangle=0.026^{+0.032}_{-0.032}$, with an intrinsic scatter $\sigma_{\rm int}=0.022^{+0.005}_{-0.005}$.
Note that the best-fit value of $a_{\rm LIV}$ is significantly less than its $1\sigma$ uncertainty bounds.
This could merely be a statistical fluke that results from the fact that the spectral lag data are quite scattered.
The best-fit $\Delta t_{\rm obs}/(1+z)$ versus $K(z)$ function (solid line; with $a_{\rm LIV}=-0.030$ and $\langle b \rangle=0.026$)
is plotted in Figure~\ref{fig1}, together with the data from GRBs. We also use the reduced $\chi^{2}$ method to constrain
the coefficients $a_{\rm LIV}$ and $\langle b \rangle$. This method also takes into account the effect of intrinsic scatter, i.e.,
\begin{equation}\label{reduced}
\chi^{2}=\sum_i \frac{\left(\frac{\Delta t_{i}}{1+z_{i}}-a_{\rm LIV}K_{i}-\langle b \rangle\right)^{2}}{\sigma_{\rm int}^{2}+\left(\frac{\sigma_{\Delta t_{i}}}{1+z_{i}}\right)^{2}}\;,
\end{equation}
where the intrinsic scatter $\sigma_{\rm int}$ is set by requiring that the reduced $\chi^{2}$ equal unity, which is widely used in Type Ia supernova cosmology
(e.g., \citealt{2012ApJ...746...85S}). Here the value of $\sigma_{\rm int}$ is 0.021 when the reduced $\chi^{2}$ is unity. We show the constraints
from the reduced $\chi^{2}$ method in Figure~\ref{fig2}(b). The best-fit values are $a_{\rm LIV}=-0.031^{+0.116}_{-0.116}$ and
$\langle b \rangle=0.026^{+0.029}_{-0.029}$, which are quite consistent with those determined from the MLE method. The corresponding
best-fit $\Delta t_{\rm obs}/(1+z)$ versus $K(z)$ function is presented in Figure~\ref{fig1} with a dashed line.

Since $a_{\rm LIV}=\Delta E'/H_0E_{\rm QG}$, for a given $\Delta E'$, $a_{\rm LIV}\rightarrow0$ gives $E_{\rm QG}\rightarrow \infty$.
Our constraints show that the slope $a_{\rm LIV}$ is consistent with $0$ within the $1\sigma$ confidence level,
and therefore there is no evidence of LIV. That is, Lorentz invariance is consistent with the data.

Marginalizing the likelihood function over the intercept parameter $\langle b \rangle$ and the intrinsic scatter $\sigma_{\rm int}$,
one can place 95\% confidence limit on the energy scale $E_{\rm QG}$ of LIV by solving the equation \citep{2006APh....25..402E}
\begin{equation}
\frac{\int_{E_{\rm QG}}^{E_{\infty}}\mathcal{L}_{\rm marg}(E)\;{\rm d}E}{\int_{0}^{E_{\infty}}\mathcal{L}_{\rm marg}(E)\;{\rm d}E}
= \frac{\int_{a_{\rm min}}^{a_{\rm LIV}}\mathcal{L}_{\rm marg}(a)\;{\rm d}a}{\int_{a_{\rm min}}^{\infty}\mathcal{L}_{\rm marg}(a)\;{\rm d}a}=0.95\;,
 \label{eq:marg}
\end{equation}
where $E_{\infty}$ represents a reference point fixing the normalization. Here we choose the Planck energy scale
as the reference point $E_{\infty}=10^{19}$ GeV, and then $a_{\rm min}=\Delta E'/H_0E_{\infty}$. Also,
$\Delta E'=E'_{h}-E'_{l}$ is the energy difference between the fixed rest-frame energy bands.
The 95\% confidence-level lower limit obtained by solving Equation~(\ref{eq:marg}) is
\begin{equation}
E_{\rm QG}\geq2.0\times10^{14}\; {\rm GeV}
\end{equation}
for the MLE method, and
\begin{equation}
E_{\rm QG}\geq2.2\times10^{14}\; {\rm GeV}
\end{equation}
for the reduced $\chi^{2}$ method.

\subsection{Test redshift variation of the LIV effect}

\begin{figure}
\vskip-0.1in
\centerline{\includegraphics[angle=0,width=0.5\textwidth]{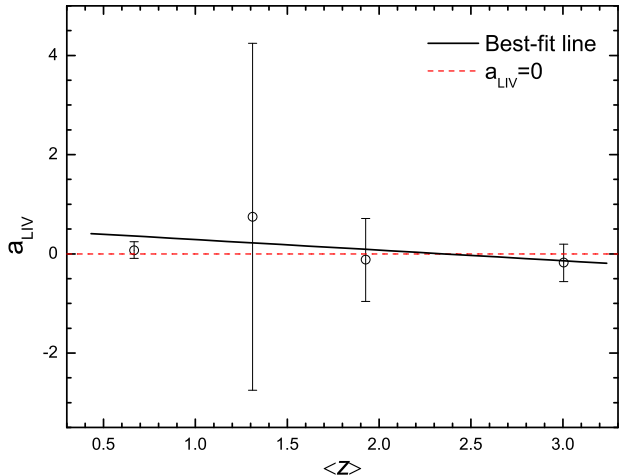}}
\vskip-0.1in
\caption{The fitted values of $a_{\rm LIV}$ against the mean redshift of GRBs.
Each data point with $1\sigma$ error bars corresponds to a group of GRBs (A, B, C and D,
from left to right). The solid and dashed lines represent the best-fit line and $a_{\rm LIV}=0$, respectively.}
\label{fig3}
\end{figure}

To test whether the LIV effect varies with redshift, we
divide the sample of 56 GRBs into four groups with redshifts
from low to high, each group containing 14 GRBs:
\begin{itemize}
  \item (A) $0.35\leq z < 0.97$, $\langle z \rangle=0.67$;
  \item (B) $0.97\leq z < 1.61$, $\langle z \rangle=1.31$;
  \item (C) $1.61\leq z \leq 2.20$, $\langle z \rangle=1.93$;
  \item (D) $2.30\leq z \leq 5.47$, $\langle z \rangle=3.00$.
\end{itemize}
For each group, we perform the same fit procedure (the MLE method) as applied to the whole GRB
sample to optimize the parameters $a_{\rm LIV}$, $\langle b \rangle$ and the intrinsic scatter $\sigma_{\rm int}$,
and calculate the mean redshift $\langle z \rangle$. The constraints on the slopes $a_{\rm LIV}$ against $\langle z \rangle$ for four groups
are shown in Figure~\ref{fig3}, together with the $1\sigma$ error bars of $a_{\rm LIV}$.
Since the data points in the second redshift bin are too much scattered, the error
bars of $a_{\rm LIV}$ are larger than those of the other three redshift bins.
A linear fit to $a_{\rm LIV}$--$\langle z \rangle$ (the solid line in Figure~\ref{fig3}) leads to
$a_{\rm LIV}=(0.501\pm0.504)-(0.213\pm0.261)\langle z \rangle$.
Clearly, the slope of $a_{\rm LIV}$ versus $\langle z \rangle$ is consistent with zero at the $1\sigma$ confidence level.
This result shows no statistically significant evidence for the redshift evolution of the LIV effect.
Moreover, because $a_{\rm LIV}$ is consistent with $a_{\rm LIV}=0$ within the $1\sigma$ confidence level for all four redshift bins,
we can conclude that there is no evidence of LIV and results that all are consistent with each other.

\section{Discussion and conclusion}
\label{sec:summary}

The spectral lags of GRBs between two fixed energy bands in the rest frame have been investigated by projecting
these two bands to the observer frame. In this work, we first use the rest-frame spectral lags of 56 GRBs to
test the possible LIV effect and its redshift dependence. This is a step forward in the investigation of LIV,
since all current studies used lags extracted in the observer frame only and the evolution of the LIV effect with
redshift has not yet been explored.

The key challenge in such time-of-flight tests, however, is distinguishing an intrinsic time delay at the source
from a possible time delay induced by the LIV effect. In order to overcome the intrinsic time delay problem,
we perform a linear fit to the spectral lags ($\Delta t_{\rm obs}$, between two fixed rest-frame energy bands in a complete sample of 56 GRBs)
versus the $K(z)$ data (it is a function of the redshift), where the slope $a_{\rm LIV}$ in the linear regression analysis
corresponds to the QG scale related to the LIV effect, and the intercept $\langle b \rangle$ represents the average effect of the intrinsic time lags
of different pulses internal to the GRB.
This technique we used was originally proposed by \cite{2006APh....25..402E}, while they focused on
the observer-frame spectral lags.

With the whole sample, we first optimize the coefficients $a_{\rm LIV}$, $\langle b \rangle$ and the intrinsic scatter
$\sigma_{\rm int}$ by maximizing the likelihood function (i.e, the MLE method). The marginalized $1\sigma$ results are
$a_{\rm LIV}=-0.030^{+0.128}_{-0.127}$, $\langle b \rangle=0.026^{+0.032}_{-0.032}$, and $\sigma_{\rm int}=0.022^{+0.005}_{-0.005}$.
Using the reduced $\chi^{2}$ method, we obtain $a_{\rm LIV}=-0.031^{+0.116}_{-0.116}$ and
$\langle b \rangle=0.026^{+0.029}_{-0.029}$. The results from these two fitting methods are in good agreement.
Since the slope $a_{\rm LIV}$ is in good agreement with $0$ within the $1\sigma$ confidence level,
there is no evidence of LIV. That is, Lorentz invariance is consistent with the data.
By marginalizing the likelihood function over $\langle b \rangle$ and $\sigma_{\rm int}$,
we obtain the 95\% confidence-level lower limit on the energy scale $E_{\rm QG}$ of LIV, yielding $E_{\rm QG}\geq0.2\times10^{15}$ GeV.
While our constraint is not the tightest, there is nonetheless merit to the result.
Thanks to our improved statistical technique and the first use of the rest-frame spectral lags,
our constraint is much more statistically significant than previous results.

By dividing the 56 GRBs into four groups according to their redshifts and fitting each group separately,
we find the best-fit values of $a_{\rm LIV}$ do not vary with the cosmological redshift, implying
there is no evidence for the reshift evolution of the LIV effect.
Moreover, because $a_{\rm LIV}$ is consistent with $a_{\rm LIV}=0$ within the $1\sigma$ confidence level for all four redshift bins,
we can conclude that Lorentz invariance is consistent with the data for all four redshift bins.
This is the first time to explore the redshift variation of the LIV effect.

\acknowledgments
We are grateful to the anonymous referee for insightful comments.
This work is partially supported by the National Basic Research Program (``973'' Program)
of China (Grant No. 2014CB845800), the National Natural Science Foundation of China
(Grant Nos. 11673068, 11603076, and 11725314), the Youth Innovation Promotion
Association (2011231 and 2017366), the Key Research Program of Frontier Sciences (QYZDB-SSW-SYS005),
the Strategic Priority Research Program ``Multi-waveband gravitational wave Universe''
(Grant No. XDB23000000) of the Chinese Academy of Sciences, and the Natural Science Foundation
of Jiangsu Province (Grant No. BK20161096).


\end{document}